\documentclass{emulateapj}

\usepackage{apjfonts}
\usepackage{graphicx}

\newcommand{\flux}{erg~s$^{-1}~$cm$^{-2}$}

\shorttitle{Multi-wavelength observations of GX 339--4}
\shortauthors{Jeroen Homan et al.}

\begin{document}


\title{Multi-wavelength observations of the 2002 outburst of GX 339--4:\\ two patterns of X-ray--optical/near-infrared behavior}


\author{Jeroen Homan\altaffilmark{1,2}, Michelle Buxton\altaffilmark{3}, Sera Markoff\altaffilmark{1}, Charles D. Bailyn \altaffilmark{3}, Elisa Nespoli\altaffilmark{2}, and Tomaso Belloni\altaffilmark{2}}

\altaffiltext{1}{Center for Space Research, MIT, Cambridge, MA 02139; homan@space.mit.edu, sera@space.mit.edu}

\altaffiltext{2}{INAF/Osservatorio Astronomico di Brera, 23807, Merate (LC), Italy; belloni@merate.mi.astro.it}

\altaffiltext{3}{Astronomy Department, Yale University, 
    New Haven, CT 06520; buxton@astro.yale.edu, bailyn@astro.yale.edu}

\begin{abstract}

We report on quasi-simultaneous {\it Rossi X-ray Timing Explorer} and
optical/near-infrared (nIR) observations of the black hole candidate
X-ray transient GX 339--4.  Our observations were made over a time
span of more than eight months in 2002 and cover the initial rise and
transition from a hard to a soft spectral state in X-rays. Two
distinct patterns of correlated X-ray--optical/nIR behavior were
found. During the hard state the optical/nIR and X-ray fluxes
correlated well, with a nIR vs.\ X-ray flux power-law slope similar
to that of the correlation found between X-ray and radio fluxes in
previous studies of GX 339--4 and other black hole binaries. As the
source went through an intermediate state the optical/nIR fluxes
decreased rapidly and once it had entered the spectrally soft state,
the optical/nIR spectrum of GX 339--4 was much bluer and the ratio of
X-ray to nIR flux was higher by a factor of more than 10 compared to
the hard state. In the spectrally soft state changes in the nIR
preceded those in the soft X-rays by more than two weeks, indicating
a disk origin of the nIR emission, and providing a measure of the
viscous time scale. A sudden onset of nIR flaring of $\sim$0.5
magnitudes on a time scale of a day was also observed during this
period. We present spectral energy distributions, including radio
data, and discuss possible sources for the optical/nIR emission. We
conclude that in the hard state this emission probably originates in
the optically thin part of a jet and that in none of the X-ray states
X-ray reprocessing is the dominant source of optical/nIR emission.
Finally, comparing the light curves from the ASM and PCA instruments,
we find that the X-ray/nIR delay depends critically on the
sensitivity of the X-ray detector, with the delay inferred from the
PCA (if present at all) being a factor of 3-6 times shorter than the
delay inferred from the ASM; this may be important in interpreting
previously reported X-ray-optical/nIR lags.

 \end{abstract}


\section{Introduction}\label{sec:intro}

Optical radiation from low-mass X-ray binaries is often attributed to
thermal emission from the outer accretion disk or companion star
\citep[see e.g.][]{vamc1995}, either as the result of internal
heating or reprocessing of X-rays from the central source. Strong
observational evidence for reprocessing in neutron star systems is
provided by the optical counterparts of type-I X-ray bursts
\citep[see e.g.][]{mccaco1979,mamioh1984}. Both their magnitude and
time lag with respect to the X-ray peak (which is on the order of
seconds) are in line with what is expected for reprocessing in the
outer disk. In black hole systems, however, the larger accretion
disks may be able to account for most of the optical radiation.
Evidence for reprocessing is not as direct; type-I X-ray bursts are
not present and often special techniques \citep[e.g.][]{obkohy2002}
are required to search for the effects of X-ray reprocessing. Based
on the correlations between the X-ray, H$_\alpha$, and optical
continuum emission, \citet{hychga2004} found evidence of reprocessing
in V404 Cyg. In XTE J1118+480 the detection of the Balmer jump
\citep{mchaga2001} argues for direct disk emission, but detection of
rapid UV, optical, and infrared variability cannot be explained by
disk emission (either direct or as the result of reprocessing). Fast
optical variability has also been observed in GX 339-4
\citep{moilch1982,moripa1983} and it has been suggested that such
variability in the optical flux of black hole binaries has to be due
to a process other than thermal emission, e.g.\  cyclotron or
synchrotron emission \citep[see
e.g.][]{fagumo1982,medifa2000,kastsp2001,hyhacu2003}.

Black hole soft X-ray transients (SXTs)
\citep{tale1995,chshli1997,mcre2003} provide an excellent opportunity
to investigate the importance of irradiation and non-thermal
processes to the optical spectra of X-ray binaries. These systems
spend most of their time in quiescence, but occasionally they show
outbursts as the result of a dramatic increase in mass accretion
rate. During these phases, SXTs brighten by many orders of magnitude
at almost all wavelengths. X-ray monitoring during outbursts has
revealed a rich behavior that is usually described in terms of
distinct phenomenological X-ray states and the transitions between
them. The nomenclature of these states has undergone many and, at
times, confusing changes in the past. Our understanding of black hole
state is still evolving and in order to avoid the risk of using a
classification scheme that may soon be outdated, we opt for a very
simplified nomenclature in this paper. We recognize only three
states, based on the slope of the spectral power-law index and the
presence/strength of band-limited noise in the power spectrum. These
states are: 1) a spectrally hard state, in which the spectrum  is
dominated by a power-law component with photon index $\sim$1.3-1.5
and the spectrum is dominated by strong ($\sim$20--40 \% rms)
band-limited (flat-topped) noise, 2) an spectrally intermediate
state, in which the power-law index varies between $\sim$1.5 and
$\sim$2.5 and the band-limited noise and QPOs increase in frequency
and decrease in strength ($\sim$10--20 \% rms) with photon index 3) a
spectrally soft state, in which the spectrum is dominated by a
thermal component, the power-law index is  $\sim$2.5 or higher, and
the power spectrum does not show strong band limited noise, but
relatively weak red noise, with or without QPOs. Our definition of
the spectrally soft state includes observations that would be
classified as steep power-law state or thermal-dominant state by
\citet{mcre2003}, but for this paper we do not distinguish between
these states, as they turn out to have very similar optical/nIR
properties. It is important to realize that the criteria summarized
above are merely guidelines. One usually needs to study an entire
outburst to identify groups of distinct behavior before   individual
observations can be classified. A full characterization of the states
in GX 339--4 is going to be addressed in \citet{hobene2004} and
\citet{behoca2004}; a full description of states is not relevant to
this paper. Here it is enough that to distinguish the hard and
intermediate states from the others. 

The observed changes in the spectral and variability properties
between the X-ray states suggests that the geometry of the inner
accretion flow evolves considerably from one state to the other. The
efficiency of irradiation of the disk (and companion star) is
probably affected by such changes in the geometry; e.g., a spherical
geometry, such as a compact corona, emits radiation more
isotropically than the (presumably) flat inner parts of a thin
accretion disk. Also, radio observations indicate that in the hard,
and intermediate states a strong outflow is often present
\citep{fe2003}, which may contribute to the infrared, optical and
X-ray fluxes \citep{mafafe2001}. Significant spectral changes are
therefore expected to occur and are, in fact, observed in the
optical/nIR \citep[see e.g.][]{moilch1985,jabaor2001b} as
the source moves from one state to another. These changes can be used
to distinguish the different mechanisms responsible for the
optical/nIR as well as to constrain models for the accretion flow
geometry in the different X-ray states.

In this paper we present (quasi-)simultaneous X-ray and optical/nIR
observations of the black hole SXT GX 339--4. This source was
discovered with $OSO$-7 in 1972 \citep{macacl1973} and an optical
counterpart was identified by \citet{dobrjo1979}. Since then it has
undergone several outbursts during which it was observed in all
states at X-ray, optical/nIR, and radio wavelengths \citep[see
references in][]{livava2001}.  It is an easy target to observe in
optical and infrared bands with an $m_V$ of 15.4--20. A mass function
of 5.8$\pm$0.5 M$_\odot$ was established by \citet{hystca2003}
strongly supporting the black hole classification of the primary. 
Optical observations of GX 339--4 in 1981
\citep{moilch1982,moripa1983} revealed fast variations and even a 20
s QPO which were interpreted by \citet{fagumo1982} as varying
cyclotron emission from a very hot gas. \citet{moilch1985} report a
sudden decline in the $V$-band magnitude as the source made a
transition from the a spectrally hard to a spectrally soft X-ray
state, after which an anti-correlation was observed between $m_V$ and
the 3--6 keV flux. They were able to fit the hard state spectral
energy distribution (SED) of the source from the infrared
($log_{10}\nu \,(Hz) \sim$14) to X-rays ($log_{10}\nu\,(Hz)\sim$19)
with a single power-law, while the soft state SED showed strong
deviations from a single power-law. On the basis of this they
suggested that different optical emission mechanisms operate in the
two X-ray states. Their hard state SED is of particular interest as it
has recently been suggested that the in the hard state the SED may be
dominated from radio to X-ray wavelengths by emission from a jet
outflow \citep{mafafe2001,manoco2003}. 

Monitoring of GX 339--4 in the hard state revealed that radio and
X-ray fluxes correlate over more than three decades of X-ray flux
over a four year time period  \citep{conofe2003}. This correlation
extends to other SXTs \citep{gafepo2003} and even AGN
\citep{mehedi2003,fakoma2003}.   Here we show that a similar
correlation is present during the 2002 hard state of GX 339--4
between optical/nIR and X-rays over a similar range in X-ray flux.
The observations discussed here were performed during the rise of an
outburst that started early April 2002
\citep{smswhe2002,beneho2002,fecotz2002,behoca2004}. Thanks to an
extensive monitoring campaign (by David Smith and co-workers) with
the {\it Rossi X-ray Timing Explorer} ({\it RXTE}\,) we were able to
follow the source in X-rays from quiescence to the peak of the
outburst. A similar optical/nIR monitoring campaign was started just
before the outburst,  which resulted in an excellent
X-ray--optical/nIR dataset of a black hole X-ray transient during the
early phases of an outburst. We give an overview of our data in \S
\ref{sec:obs}  and present our analysis and results in \S
\ref{sec:res}. Possible interpretations of the observed behavior are
discussed in \S \ref{sec:disc}. Conclusions and a summary are
presented in \S \ref{sec:sum}.

\section{Observations and Data Reduction} \label{sec:obs}

\subsection{Optical and Infrared Photometry}

Optical photometry of GX 339--4 was taken between UT 2002 February 10
-- May 29 (MJD 52315--52423) on a daily basis when possible. 
Near-infrared photometry was obtained between UT 2002 January 21 --
September 24 (MJD 52295--52541).  Observations in the optical ceased
earlier than near-infrared due to the failure of the optical CCD. 
Observations were made with the YALO\footnote{Yale, AURA, Lisbon,
Ohio consortium, http://www.astronomy.ohio-state.edu/YALO/. YALO is
now superseded by SMARTS (Small and Medium Aperture Research
Telescope System), http://www.astro.yale.edu/smarts/, which currently
uses ANDICAM on the 1.3m CTIO telescope.} 1.0m telescope at CTIO
using the
ANDICAM\footnote{http://www.astronomy.ohio-state.edu/ANDICAM/}
camera. ANDICAM takes simultaneous optical and infrared images over a
variety of band-passes.  In our observations we used Johnson $V$-,
$I$- and CIT $H$-band filters.

Exposure times for the optical images were 240 sec in $V$ and 200 sec
in $I$.  Optical images were bias subtracted and flat-fielded using
{\tt CCDPROC} in {\tt IRAF}.  Photometry was performed using {\tt
DAOPHOT} in {\tt IRAF}.    In the $H$-band we obtained nine images of
50 sec each.  Each set of nine images were reduced using an in-house
{\tt IRAF} script which flat-fields, sky subtracts, shifts and
combines each set of images.  Photometry of the combined images was
performed using {\tt DAOPHOT} in {\tt IRAF}.  The errors shown on the
data points are those obtained with {\tt DAOPHOT}.  The photometry
was calibrated using the standard stars T PheD
\citep[optical,][]{la1992} and P9103\footnote{Persson faint infrared
standards,\\
http://www.ctio.noao.edu/instruments/ir\_instruments/ir\_standards/hst.html}
(infrared).

\subsection{X-ray observations}

The X-ray observations were performed with the Proportional Counter
Array \citep[PCA;][]{zhgija1993,jaswgi1996} and the High Energy X-ray
Timing Experiment \citep[HEXTE;][]{grblhe1996,roblgr1998} on-board
the {\it Rossi X-ray Timing Explorer} \citep[{\it
RXTE};][]{brrosw1993}. We analyzed data taken between UT 2001 March
6   (MJD 51974) and UT  2002 October 13 (MJD 52560), corresponding to
a total of 137 pointed observations (excluding one which was only 48
seconds long). The exposure times for the individual observations
range from 400 to 18000 s, with the observations before MJD 52360 all
being shorter than 1300 s. Only data from Proportional Counter Unit 2
(PCU2) were used for the analysis in this paper, as it is the only
detector that was operational during all observations and is the best
calibrated detector of the 5 PCUs. From HEXTE we only used cluster A
data.

PCA spectra were produced from the {\tt standard 2} mode data, which
has a time resolution of 16 s and covers the 2--60 keV range with 129
channels. HEXTE spectra were created from the standard mode {\tt
Archive} data which has a nominal 32 s time resolution and covers
the 10--250 keV range with 64 channels. Spectra were produced for each
observation using FTOOLS V5.2. All spectra were background subtracted
and dead-time corrected. A systematic error of 0.6\% was added to the
PCA spectra, which is a common practice.
The  PCA (3--25 keV) and HEXTE (20--150 keV) spectra of each
observation were fit simultaneously in {\tt XSPEC}
\citep[V11.2]{ar1996} using an overall normalizing constant that was
allowed to float for cross-calibration purposes. We used the
following combination of models: a (cut-off) power-law ({\tt
cutoffpl} or {\tt powerlaw}), a multi-temperature thermal model \citep[{\tt diskbb} - hereafter referred to as the 'disk' component, see e.g.][]{miinko1984}, a
relativistically smeared emission line around 6.5 keV ({\tt laor}), a
smeared absorption edge ({\tt smedge}), and an absorption component
({\tt phabs}). We also added an edge ({\tt edge}) at 4.79 keV with
optical depth $\tau<0.2$ to correct for an instrumental Xe edge at
this energy in the PCA spectra.  $N_H$ was fixed to a value of
$5\times10^{21}$  atoms cm$^{-2}$ \citep{kokuch2000}. This model led,
on average, to values of $\chi^2_{red}$ that were slightly smaller
than 1. Note that in the early phase of the outburst the disk
component, the emission line and the smeared edge were not
significantly detected and therefore removed from our fit model.
Also, when the source was not significantly detected with HEXTE, fits
were made to the PCA spectrum only. Unabsorbed fluxes were measured
in the 3--100 keV range.  If no fits were made to the HEXTE spectra,
the fit to the PCA spectrum was extrapolated. A more detailed
description of the spectral (and variability) analysis and a complete
overview of the fit results of the whole outburst will be presented
elsewhere.

\section{Results} \label{sec:res}

\subsection{X-ray light curves}

\begin{figure*}[t]
\centerline{\includegraphics[width=11cm]{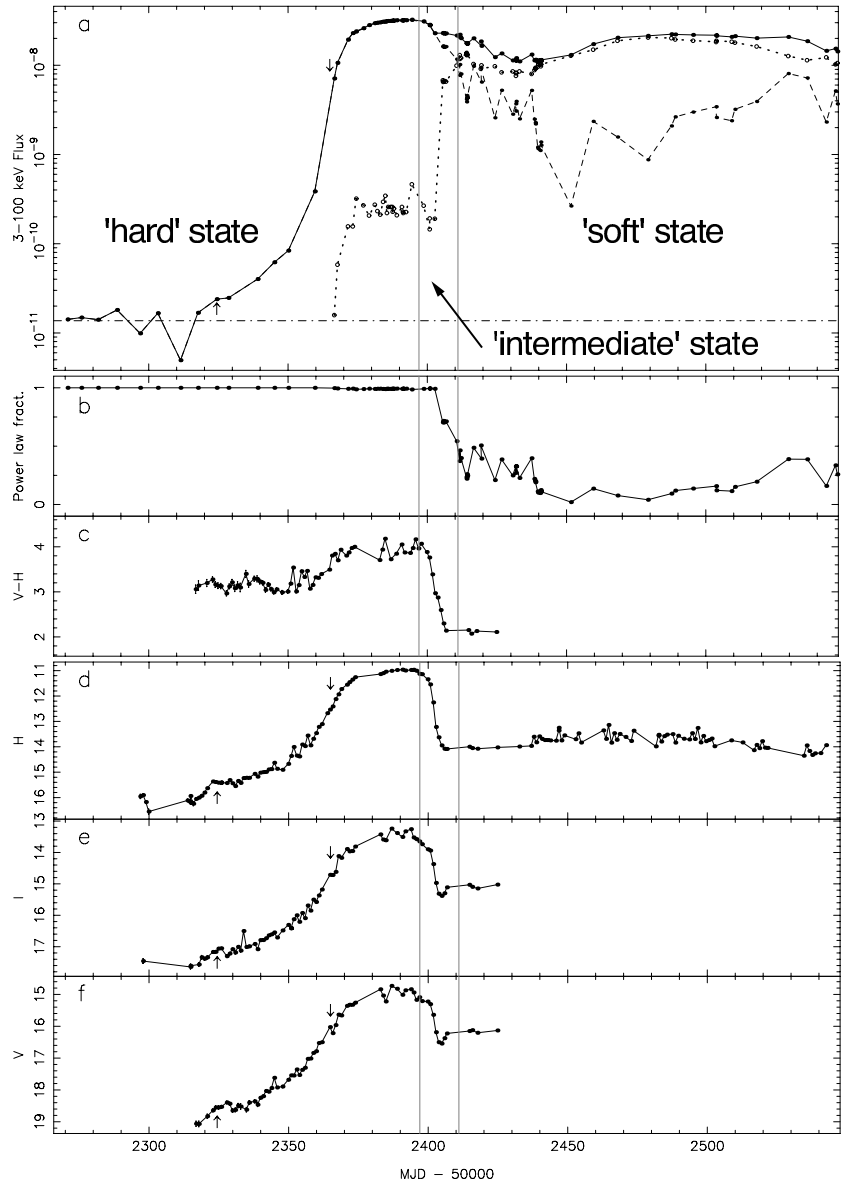}}

\caption{{\it RXTE}/PCA X-ray and YALO optical/nIR light curves of GX
339--4. (a) Total  flux (solid line), power-law flux (dashed line),
and disk flux (dotted line, open circles). All fluxes are in the
3--100 keV band and are unabsorbed. The dashed-dotted line indicates
the average 3--100 keV flux in the $\sim$55 days before the first
significant observation in the X-rays (indicated by the upward
arrow). The downward arrow indicates the date of the first detection
of the source with the {\it RXTE}/ASM. The gray vertical lines show
the approximate times of the state transitions in the X-ray band -
the states are indicated as well; (b) The fractional contribution of
the power-law component to the 3--100 keV flux; (c) {\it V-H} color;
(d--f) {\it H}, {\it I}, and {\it V} magnitudes. \label{fig:curves}}
\end{figure*}

In Figure \ref{fig:curves}a we show three X-ray light curves of GX
339--4 during the first half of its 2002/2003 outburst: the total
flux (solid line), the flux of the (cut-off) power-law (dashed line),
and the disk black-body flux (dotted line), all in the 3--100 keV
band. The first significant detection ($>3\sigma$) of the source with
the {\it RXTE}/PCA was made on MJD 52324 (marked by the upward
arrow), although the outburst likely started before this date. For
comparison we have also marked the date of the first detection with
the {\it RXTE}/ASM, on MJD 52365, by the downward arrow. Had the
source not been monitored with the {\it RXTE}/PCA we would not only
have missed out on more than 40 days of valuable X-ray data, but we
would also have miscalculated the true delay between the X-ray and
optical/nIR rise. Extrapolation of the light curve around the time of
the first ASM detection back to zero flux (based either on PCA or ASM
data)  resulted in an earlier start (15--25 days earlier), but still
considerably later than the first PCA detection. The presence of an
initial slow rise, as observed with the PCA, shows that the start
times obtained from such extrapolations should be interpreted with
some care. For the same reason we do not use the first significant
PCA observations to obtain a 'improved' start time; we simply do not
know the shape of the light curve below our detection limit. 

The dashed-dotted horizontal line Figure \ref{fig:curves}a indicates
the average flux measured in the 8 observations prior to MJD 52324.
This flux level (a significant detection of $\sim$1.4 x 10$^{-11}$
\flux) is about a factor of 10 higher than the lowest values observed
during quiescence  \citep[in September 2000 by][]{conofe2003}, but
probably represents the confusion limit for the PCA (R. Remillard,
private communication). If so, this   indicates that we may have
missed a considerable part of the rise due to the limited sensitivity
of the PCA compared to the X-ray imaging instruments (BeppoSAX and
XMM-Newton) that were used for the observations in quiescence. The
rise of the outburst was initially slow, with an increase in 25 days
by a factor of $\sim$6 with respect to the level in the 55 days
before MJD 52324. The next 30 days saw a dramatic increase by a
factor of more than 350 in flux. After MJD 52380 the rate of increase
slowed down considerably with the source entering a broad maximum for
about 20 days during which it reached a maximum flux of 3.25 x
10$^{-8}$ \flux.

Before MJD 52405 the contribution of the disk component to the X-ray
spectrum was very small, with the power-law component contributing
close to 100\% of the 3--100 keV flux, as can be seen in Figure
\ref{fig:curves}b.   No disk component was detected before MJD 52366,
after which we measured values that were typically $\sim$1\% of the
total flux. This changed dramatically on MJD 52405 when the disk flux
had increased by a factor of more than 35 compared to the previous
observation, three days earlier. This day also marked a change of
behavior in the power-law flux which, until then, had shown only
smooth day-to-day variations, but suddenly showed erratic changes on
top of a steady decline. Although the power-law fraction
(Fig.~\ref{fig:curves}b) suggests that the X-ray spectrum did not
change considerably before MJD 52405, X-ray color curves
\citep{behoca2004,hobene2004} indicate that spectral changes already
started before MJD 52400. It is interesting to see that, like the
power-law flux, the disk flux showed a decline for about 25 days
after it reached a local maximum on MJD 52414. After MJD 52438 the
source entered a long period in which the disk component contributed
more than 75\% of the 3--100 keV flux, showing a slow increase
followed by a similarly slow decrease. During this period the
power-law flux showed large variations.  It should be noted, however,
that it was in general hard to constrain the high energy part of the
spectrum in this part of the outburst, because of the low number of
counts.  

\subsubsection{X-ray states}

Based on the definitions of the X-ray states given in
\S\ref{sec:intro} we can attempt to (preliminary) classify the X-ray
observations.  The period until MJD 52398 can be classified as a hard
state; the power spectrum was dominated by a power law component with
an index of $\sim$1.3--1.4. As the luminosity increased the power-law
component showed minor steepening. This steepening accelerated around
MJD 52398, which is the day where we put the transition to the 
intermediate state. In addition to the rapid steepening of the
power-law component to an index of $\sim$2.5 we also observed
accelerated changes in the QPO frequencies \citep{behoca2004} during
the intermediate state.  Note that the exact date of the transition
from the hard state to intermediate state is to some extent arbitrary
as many of the properties show a smooth evolution across the
transition. However, the date of MJD 52398 is chosen also because it
marks a clear change in the optical and nIR properties of the source
(see \S \ref{subsec:oir} and \S \ref{ref:corr}) and in the
hardness-intensity diagram \citep{behoca2004}. The intermediate state
ended on MJD 52410. On that day the power spectrum showed strong
band-limited noise, while on the next day (and some of the following
days) we detected red noise and type B QPOs \citep{behoca2004}. Note
that during the  intermediate state, on MJD 52402, the disk flux
showed a dramatic increase, without substantially changing or
interrupting the evolution of the power-law index and type C QPO
frequencies \citep[see][for a definition if QPO
types]{wihova1999,resomu2002}. From MJD 52411 onward the source
showed several types of power spectra, but the spectra were in
general much softer (see Fig.\,\ref{fig:curves}b) and the variability weaker than before that time.
We classify all these observations as the spectrally soft state,
ignoring for the moment the significant changes in the variability
properties within that state. The approximate times of  the
hard/intermediate state and intermediate/soft state transitions are
indicated by the vertical lines in Fig.~\ref{fig:curves}.

\begin{figure*}[t]\centerline{\includegraphics[angle=-90,width=8cm]{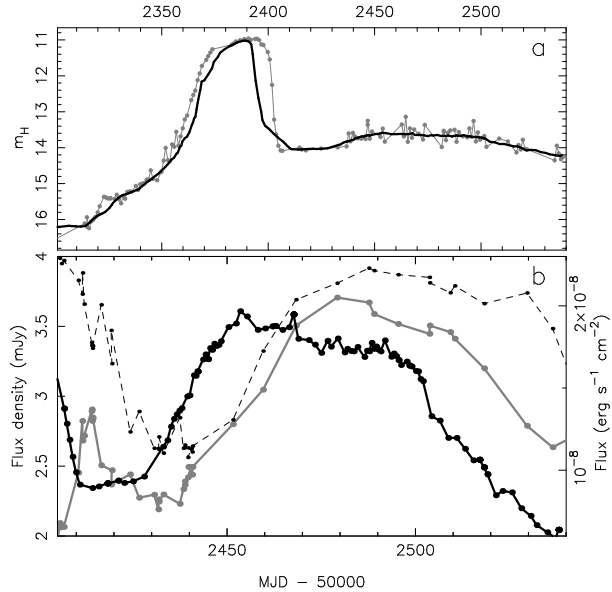}}
\caption{(a) Original non-rebinned {\it H}-band light curve (gray points/line) and its 20-day running average (thick solid line); (b) A comparison of the 20-day running average of the {\it H}-band flux density (thick black line) with the disk (thick gray line) and  total (dashed line) X-ray flux in the spectrally soft state.   \label{fig:flaring}}
\end{figure*}

\subsection{Optical/nIR light curves}\label{subsec:oir}

The outburst in the optical/nIR (Fig.\,\ref{fig:curves}d--f) started
around MJD 52317, a week before our first significant detection in
X-rays.  We note, however, that this start is also consistent with
zero delay between X-rays and the optical/nIR (see also \S
\ref{ref:corr}), as the first PCA detection does not necessarily
represent the true start of the outburst in X-rays. Between MJD
52296--52300 the $H$-band light curve indicates the possibility of a
dip before the outburst.  It is difficult to place any constraints on
this, however, due to the lack of data during this time.  After  MJD
52317, when the source was in the hard state, the optical/nIR light
curves all showed a profile similar to that of the X-ray flux; an
initially slow increase followed by a faster rise that evolved into a
broad maximum. While the optical/nIR color (Fig.\ref{fig:curves}c)
remained at the same level for the first 35 days of the outburst, a
clear reddening was observed around MJD 52360.

Similar to the X rays, the optical/nIR fluxes started decreasing as
the source entered the intermediate state.  After an initially slow
decline the optical/nIR light curves decreased dramatically around
52400 by more than 3 mag in $H$, more than 2 mag in $I$, and almost 2
mag in $V$, all within 5 days. During this short phase the {\it V-H}
dropped from 4 to 2, indicating the spectrum became much bluer. By
comparison, the decay in X-rays was much slower than in the
optical/nIR, with a time scale of $\sim$30--40 days. After the fast
decline a short dip ($\sim$3--4 days) was observed in the {\it I}-
and {\it V}- bands, but not in the {\it H}-band. After this dip the
optical/nIR remained more or less constant.

A few days before the disk component started to dominate the X-ray
spectrum two changes occurred in the $H$-band light curve (see
Fig.~\ref{fig:flaring}a).   First, the source brightened by about
half a magnitude and, second,  there was a sudden onset of
variability. The variations had an amplitude of  $\Delta m_H\sim0.5$
and a typical time scale of a day (although our sampling does not
exclude faster variations). They were were not observed in one or
more of our three reference stars, hence they are intrinsic to GX
339--4. Also, the apparent lack of variability in $H$-band between
MJD 52410--52435 can not be explained by the sparser sampling of the
source during that period.  By randomly removing points from the
light curve in order to obtain a similar sampling as that between MJD
52410--52435, we were not able to hide the large variation observed
after MJD 52435. Note that while the onset of flaring nearly
coincided with the disk component becoming the dominant X-ray
spectral feature, the flaring remained present when the power-law
component became more prominent again, at the end of our data set.
Unfortunately no observations were made in the {\it V}- and {\it
I}-bands during the flaring period, so it is not known whether
similar variations were also present in the optical.

To get a better idea of the long-term changes in the {\it H}-band
during the flaring period we calculated a 20 day running average of
the total {\it H}-band light curve (running averages with 15--25 day
intervals produced similar results). In addition to the first peak
around MJD 52390, the smoothed curve (Fig.~\ref{fig:flaring}a) shows
a second broad maximum that peaks around MJD 52450. In
Fig.~\ref{fig:flaring}b we compare this second $H$-band maximum with
the total and disk X-ray fluxes from the same period. Note the
change from a magnitude scale in Fig.~\ref{fig:flaring}a to a
(linear) flux density scale in Fig.~\ref{fig:flaring}b. In X-rays the
source reached a peak well after that in the {\it H}-band
with a delay of $\sim$15--20 days.  By actually
shifting the light curves in time with respect to each other, we find
that they overlap best with a shift of about 18 days.

\begin{figure*}[t]
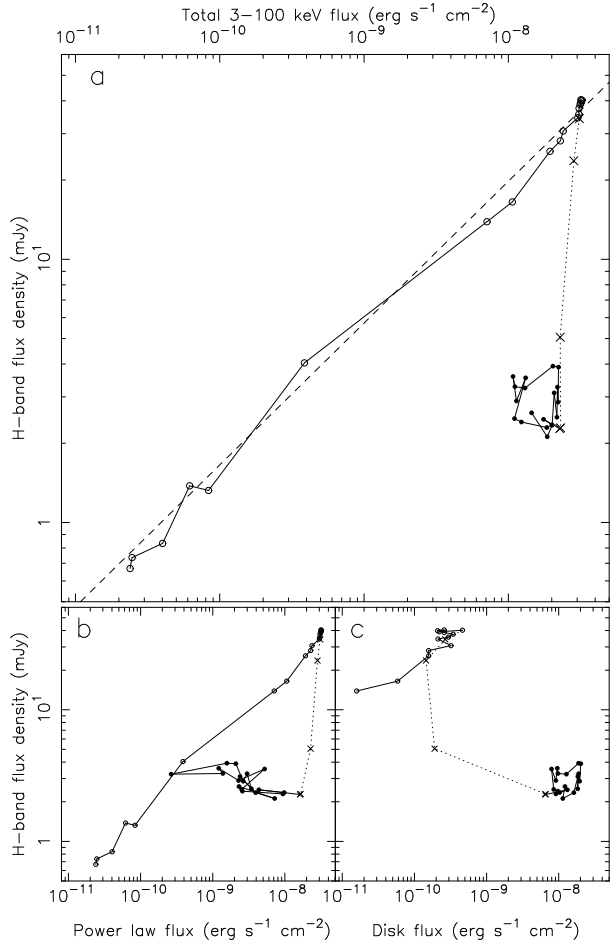

\centerline{\includegraphics[angle=-90,width=8cm]{f3a.eps}}
\centerline{\includegraphics[angle=-90,width=8cm]{f3bc.eps}}
\caption{$H$-band flux density versus (a) total flux; (b) power-law flux; and (c) disk flux. Symbols are used to indicate X-ray state: hard state (open circles), intermediate state (crosses) and spectrally soft state (filled circles). The dashed line in panel $a$ shows the best power-law fit to the hard state data (open circles), with an index of 0.53$\pm$0.02\label{fig:flux-flux}}
\end{figure*}

\subsection{X-ray--optical/nIR correlations} \label{ref:corr}

From Figure \ref{fig:curves} it is clear that in the hard state the
profiles of the optical/nIR light curves are similar to that of the
power-law flux light curve. In the spectrally soft state, however,
the nIR light curve is much more similar to that of the disk flux,
with a much lower IR/X-ray ratio than in the hard state.  The
intermediate state seems to represent the transition between these
two types of behavior. To illustrate this, we show a plot of {\it
H}-band flux density versus total X-ray flux in Figure
\ref{fig:flux-flux}a. The {\it H}-band magnitudes were converted back
to flux density units, only using data points from the X-ray and {\it
H}-band light curves that were taken within less than 2 days from
each other. The open circles in Figure \ref{fig:flux-flux}a are from
the (hard state) rise of the outburst until $\sim$MJD 52394
(corresponding to the peaks in X-rays and the {\it H}-band light
curves). As can be seen, the X-ray and {\it H}-band fluxes in this
state are strongly correlated, with a correlation coefficient of
0.99, although some scatter is present. Fitting these points with a
power-law yields $\alpha$ = 0.53$\pm$0.02 where $F_H \propto F_X
^{\alpha}$.  The relation seems to steepen somewhat toward higher
X-ray fluxes. Fits with a power-law to the points with an X-ray flux
smaller and larger than $10^{-9}$ \flux\ give indices of 0.65$\pm$0.03
and 0.73$\pm$0.03, respectively, which, unlike the overall relation,
are consistent with the slope of radio/X-ray relations found by
\citet{conofe2003} and \citet{gafepo2003}. The fact that both are
steeper than the overall correlation is due to the relative
displacement of the two groups, with the low X-ray flux group having
a relatively higher {\it H}-band flux density. Such a displacement
could in principle be explained if there is another underlying
component to the $H$-band flux, but not for the X-ray flux, so that
at low light levels, the $H$-band levels off whereas the X-ray does
not. Note that errors on the X-ray fluxes obtained from XSPEC are
hard to estimate, especially when multi-component models are used.
Power-law slopes measured using a wide range in errors were all
consistent with each other.

After the peak of the outburst, when it entered the intermediate
state,  the source started following a path that clearly deviated
from the one traced out in the hard state. This path is indicated by
the dotted line and the crosses in Figure \ref{fig:flux-flux}a. 
During this period the {\it H}-band flux dropped by a factor of
$\sim$17, while the X-ray flux dropped by a factor of only $\sim$1.4.
In the spectrally soft state the source traced
out a clockwise pattern, which probably corresponded to the
$\sim$15--20 day delay measured between the smoothed {\it H}-band
curve and X-ray curve. 

In  Fig.~\ref{fig:flux-flux}b and \ref{fig:flux-flux}c we show
correlations between the {\it H}-band flux density, and the power-law
and disk flux, respectively. Comparing these panels to panel (a)
confirms that during the rise of the outburst the {\it H}-band flux
correlates well with the power-law component and that the circular
pattern in the spectrally soft state is indeed due to a delay
between the disk flux and {\it H}-band flux density. Plots of {\it
V}-band and {\it I}-band flux densities versus total X-ray flux (not
shown) are similar to those for the {\it H}-band flux density . Both
correlate well during the hard state and can be fitted with a
(slightly flatter) power-law with indices of 0.48$\pm$0.02 for the
{\it I}-band and 0.44$\pm$0.03 for the {\it V}-band.

\begin{figure*}[t]
\centerline{\includegraphics[angle=-90,width=8cm]{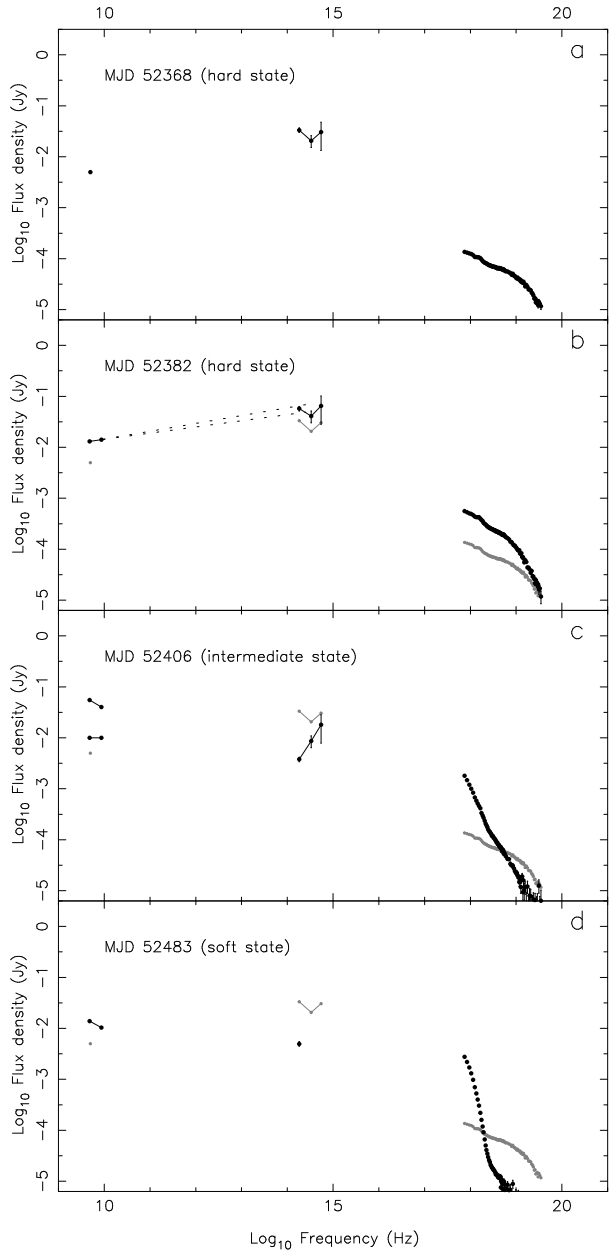}}
\caption{Spectral energy distributions (SEDs) of four selected observations. The SED of panel $a$\label{fig:sed} is also shown in gray in the other three panels. For panel ($a$) only one radio data-point was available and for panel ($d$) no {\it I} and {\it V}-band data were available. The two dotted lines in panel ($b$) indicate the range of values expected on the basis of an extrapolation of the spectral slope in the radio. The two radio datasets in panel ($c$) represent the start and peak of the radio flare observed by \citet{gacofe2004}. Errors are plotted for all data-points, but for most points they are  smaller than the plot symbols.\label{fig:seds}}
\end{figure*}

\subsection{Spectral Energy Distributions}

To relate the (spectral) changes in the optical/nIR, X-ray and radio
bands to each other, we have created several spectral energy
distributions (SEDs). The radio flux densities were taken and
estimated from \citet{fecotz2002} and \citet{gacofe2004} (also
E.~Gallo and S.~Corbel, private communication). Our optical/nIR
magnitudes were converted to flux densities \citep{zo1990} and
dereddened assuming an $N_H$ of $5\pm1 \times 10^{21}$ atoms
cm$^{-2}$ \citep{kokuch2000}. The $N_H$ was converted
\citep{prsc1995} to a reddening E(B-V) of 0.94, which was in turn
converted into dereddening factors using the values found in
\citet{caclma1989}. These factors were 1.67 ({\it H}\,), 3.63 ({\it
I}\,), and 14.78 ({\it V}\,). Our X-ray spectra were deconvolved
using the model fit (with $N_H$ set to zero) and then converted to
flux densities \citep{zo1990}. Figure \ref{fig:seds} shows four SEDs:
two from the hard state (MJD 52368, with X-ray data from MJD 52367,
and MJD 52382), one from the intermediate state (MJD 52406; radio
data from MJD 52408), and one from the spectrally soft state (MJD
52483; X-ray data from MJD 52487). For comparison we plot the SED of
MJD 52368 also in the other three panels (gray dots). Note that the
errors on the optical/nIR points are correlated through the
uncertainty on $N_H$ (which dominates the error) and should therefore
have the same direction.

\begin{figure*}
\centerline{\includegraphics[angle=-90,width=8cm]{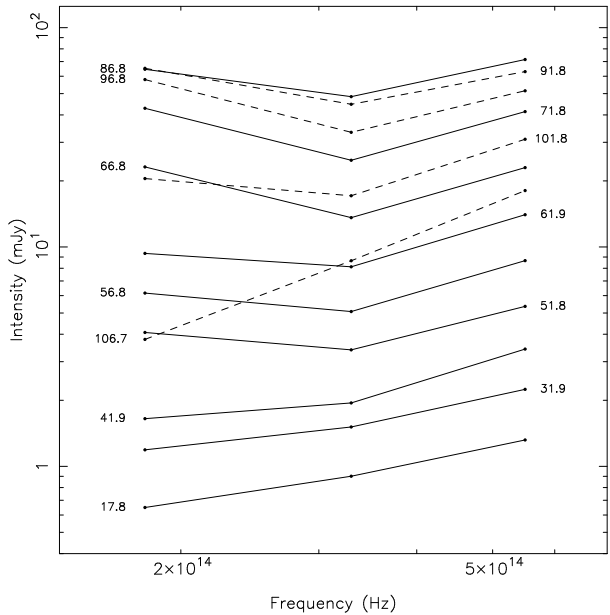}}
\caption{Optical/nIR spectral energy distributions (SEDs) during the rise in the hard state and transition to the spectrally soft state. Solid lines are used for observations before the peak in the {\it H}-band and dashed lines for those after. Dates of the SEDs alternate between left and right to avoid overlapping and are MJD$-$52300.   \label{fig:sed_evol}}
\end{figure*}

The first two SEDs  (Fig. \ref{fig:seds}a,b) show that the
brightening of the X-rays and optical/nIR in the hard state is
accompanied by a brightening of the radio by a similar factor. The
optical/nIR part in these hard state SEDs has a shape (see also
Fig.~\ref{fig:sed_evol}) that is very similar to the turnover
observed by \citet{cofe2002} in GX 339--4 during its 1981 and 1997
outbursts.  The slopes of power-laws that connect the radio and {\it
H}-band points for the first and second SED are 0.18$\pm$0.01 and
0.15$\pm$0.01, respectively. These slopes are very similar to those
of the optically thick/self absorbed radio spectra of black holes
sources in their hard state \citep{fe2001b,fehjti2001}. This
association is supported by the fact that the $H$-band point of the
second SED lies close to the extrapolation of the radio spectrum,
which has a slope of 0.14$\pm$0.02 (the full range is indicated by
the two dotted lines).  \citet{cofe2002} suggested that the near-infrared lies above the optically thin break of the compact jet synchrotron spectrum. Although such a break cannot be seen in our data, the negative spectrum slope between the $H$ and $I$-bands, suggests this to be the case in our hard state observations as well.

The intermediate state SED (Fig. \ref{fig:seds}c) shows a clear
steepening of the X-ray spectrum and the disappearance of the
turnover in the optical/nIR.  Two sets of radio points are shown in
the intermediate state SED. The lower one represents the base level
at the start of the radio observation, indicating a small decrease
compared to the second hard state set, and the upper one the peak of
a radio flare that began two hours after the start of the observation
and lasted for at least six hours \citep{gacofe2004}. A clear change
in the spectral index can be seen, indicating a change to more
optically thin emission as the source brightened. The radio ejecta
associated with this flare were followed for more than four months by
\citet{gacofe2004}. For the soft state SED (Fig. \ref{fig:seds}d) no
{\it I}- and {\it V}-band data were available. The X-ray spectrum was
even steeper than in the intermediate state but the {\it H}-band flux
was of a similar value. The source was still detected at radio
wavelengths at a flux density significantly higher than in the first
hard state observation. The radio spectral index and imaging
observations  \citep{gacofe2004} show that this emission was from an
extended rather than a compact source.  

In Figure \ref{fig:sed_evol} we show the evolution of the optical/nIR
part of the SED in more detail. The solid lines are from the rise
until the peak in the optical/nIR, the dashed lines from the decay. 
The {\it H}-band not only rises faster than the {\it I}- and {\it
V}-bands, but also decays much faster than those bands. The slope
between the {\it V} and {\it I} bands stayed surprisingly constant. 
Judging from similar the profiles of the three optical/nIR bands one
would think they were produced by the same component and that the
spectral reddening seen in the {\it H}-band would also be observed
in  the {\it I/V} part of the spectrum. However, this is not clearly
seen. Possible explanations for this will be discussed in \S
\ref{sec:hard_state}

\section{Discussion} \label{sec:disc}

During the first $\sim$240 days of its 2002/2003 outburst GX 339--4
was observed in three X-ray states: a hard state (before MJD 52398),
an intermediate state (MJD 52398--52410) and a spectrally soft state 
(after MJD 52411). In the hard state a clear correlation between the
optical/nIR and X-ray fluxes was observed with no measurable lag
between the X-rays and optical/nIR. In the intermediate state the
ratio of optical/nIR and X-ray fluxes decreased  by more than a
factor of 5--10 compared to the hard state. This ratio remained more
or less constant in the spectrally soft state, where long-term
changes in the X-rays lag those in the nIR by $\sim$20 days. This
difference in behavior suggests that the optical/nIR emission in the
hard state had a very different origin than in the latter two states
and that the intermediate state represents a transitional state. 
This is supported by the changes in the optical/nIR SEDs.

\subsection{Comparison to previous outbursts and other sources}\label{sec:comp} 

X-ray--optical/nIR behavior similar to that reported in \S
\ref{sec:res} has been observed before in GX 339--4. During the
hard-to-soft transition of the 1981 outburst \citet{moilch1985}
observed a fast ($<$8 days) decline in the  $V$-band intensity by a
factor of 2.5--3, while the 3--6 keV {\it Hakucho} count rate
increased simultaneously by a similar factor. For comparison,
examining a light curve  of our 3--6 keV {\it RXTE}/PCA count rate
reveals an increase by a factor of $\sim$1.6 during a period in which
the $V$-band flux dropped by a factor of $\sim$2.3.
\citet{moilch1985} were able to fit the optical/nIR and X-ray parts
of their hard state SED with a single power-law with index $-0.58$. 
None of our observations are consistent with this, not even when
using the much lower $A_V$ of \citet{moilch1985}.  We do, however,
observe a similar dramatic change between the SEDs of the hard and
soft spectral state. Our hard state SEDs are quite similar
to those in \citet{cofe2002} which show a clear turnover in the
optical/nIR range.

In the last few years long-term optical/nIR campaigns have been
undertaken for the two  black hole X-ray transients XTE J1859+226
\citep{sacagi2001,zusaca2002} and XTE J1550--564 \citep[1998/1999 and
2000 outbursts;][]{sacadu1999,jabaor1999,jabaor2001b,jabaor2001a}. 
In most of these cases the optical/nIR data were only compared to
{\it RXTE}/ASM and not to {\it RXTE}/PCA data. The optical/nIR light
curve of XTE J1859+226 peaked a few days after the {\it RXTE}/ASM
light curve, as opposed to GX 339--4 where the optical/nIR peaked
about 15 days before a (local) maximum in the ASM light curve (around
MJD 52410, not visible as a peak in the {\it RXTE}/PCA light curve).
{\it RXTE}/PCA observations show that near the end of the X-ray
coverage of XTE J1859+226 the source made a transition to a harder
spectral state.  We analyzed four of the five observations after this
transition and find that the energy spectra  are well fitted with
power-laws with indices between 1.7 and 2.3, suggesting a
transitional phase to the hard state. The transition to the harder
state occurred between MJD 51587--51608 and does not clearly stand
out in the optical/nIR. XTE J1859+226 did show several mini-outbursts
in optical/nIR. However, they all occurred when the X-ray
observations had ended and the source had presumably returned to the
hard state.

\begin{figure*}
\centerline{\hbox{\includegraphics[angle=-90,width=8cm]{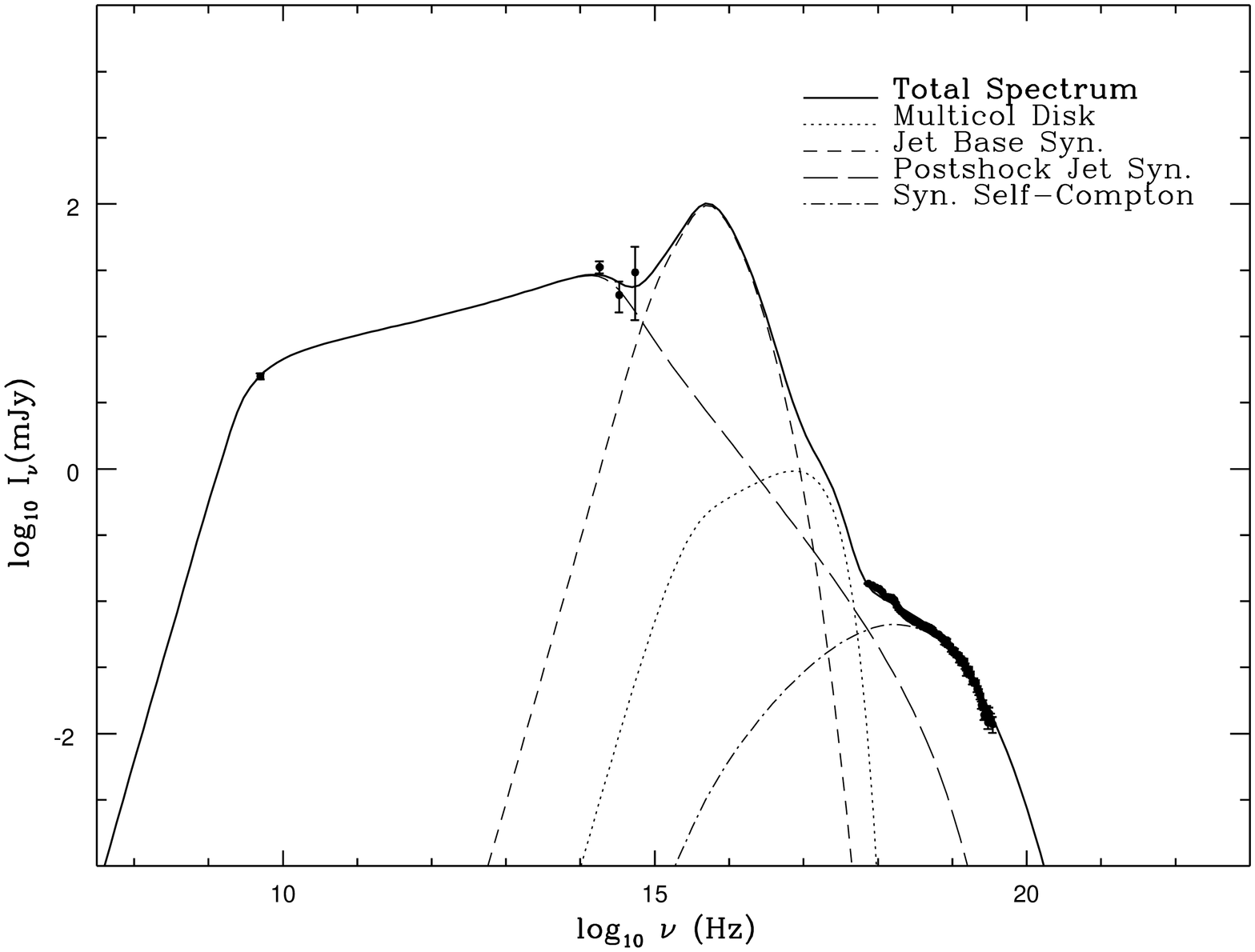}\includegraphics[angle=-90,width=8cm]{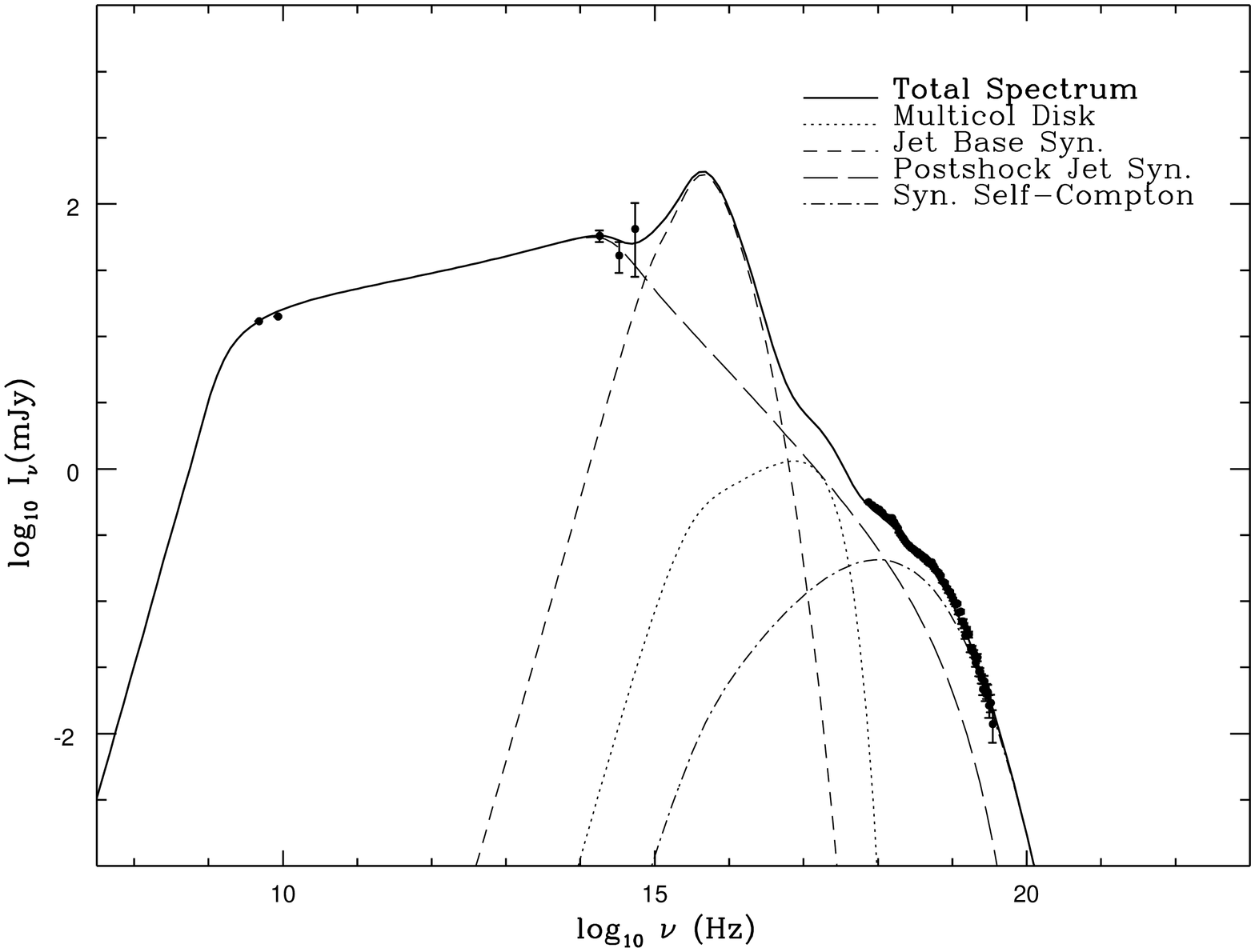}}}
\caption{Representative fits for the extreme case of entirely
jet-dominated models for the two hard state SEDs shown in Fig.~4.  The
jet plasma is assumed to enter the jet at its base, near the central
object, where the high densities and magnetic fields are conducive for
significant synchrotron (short dashed line) and synchrotron
self-Compton (dot-dashed line) emission.  This plasma must undergo
extreme cooling or some other decoupling from the outer jet (long
dashed line), which produces the radio through nIR emission via
self-absorbed synchrotron radiation, in order to show a dip around the
I-band.  A fraction of the particles (for bright luminosities as shown
here $\sim 40\%$) are accelerated in the outer jet (or else the power
requirements are too high, see \citealt{manoco2003}), resulting in a
power-law tail contributing synchrotron emission up through the
X-rays.  These models are not convolved with the detector response
matrices and do not include line emission or reflection components,
and therefore represent the maximum contribution from jets to the hard
state SED.
\label{fig:jetmodel}}
\end{figure*}

After reaching its peak, the {\it H}-band light curve of the 2000
outburst of XTE J1550--564 showed a strong decrease by $\sim$1.5
magnitudes within $\sim$8 days \citep{jabaor2001b}. This decrease
started when the source made a transition from the hard state to the
intermediate state \citep{rocoto2003}. Unlike GX 339--4, in XTE
J1550--564 a similar decrease is not observed in the {\it I}- and
{\it V}-bands. On the other hand, the return of XTE J1550--564 to the
hard state shows up as an increase in all three optical/nIR bands,
similar to what was observed in GX 339--4 when it returned to the
hard state at the end of the 2002/2003 outburst (M.~Buxton, in prep.)
and more recently in 4U 1543--47 at the end of its 2002 outburst
\citep{buba2004}. This suggests that the hard state can be associated
with an enhancement of the nIR and, to a lesser extent, the
optical. In this respect, it is important  to distinguish between the
hard state, where the hard component completely dominates, and the
other three states which do show a hard component, but a (much) weaker
and steeper one.  Observations of XTE J1550--564 during the second
part of its 1998/1999 outburst suggest an anti-correlation between
the optical/nIR and the strength of the hard spectral component in
the spectrally soft state \citep{jabaor2001a},
possibly indicating that the hard component in these states has a
different origin that the one in the hard state, as has already been
suggested before \citep[see, e.g.,][]{co1999}.  

\subsection{Origin of the optical/nIR emission}

There are several physical components to an X-ray binary, many of
which are theorized to produce significant optical/nIR emission: the
accretion disk, the secondary star, a jet \citep[e.g.][]{mafafe2001},
and/or a magnetically dominated compact corona
\citep[e.g.][]{medifa2000}. In the first two components the
optical/nIR can either be the results of internal or external heating
by X rays, depending upon the state of the system and exact
parameters. \citet{shfech2001} and \citet{chmigo2002} showed that the
companion star in GX 339--4 is most likely an evolved low-mass star
with little ($<$20--30\%) contribution in the near-infrared in
quiescence. Assuming a black hole mass of 5.8 $M_\odot$, a secondary
mass of 0.52 $M_\odot$, and an orbital period of 1.75 days
\citep{hystca2003}, which gives a binary separation of
$\sim7.8\cdot10^{11}$ cm, irradiation of the secondary by an X-ray
source with a luminosity of $1.5\cdot10^{38}$ erg s$^{-1}$ is
unlikely to increase the optical/nIR flux by more than a factor 10,
much less than the factor of a few hundred required to explain the
observed optical/nIR increase. This basically rules out the secondary
as the dominant source of optical/nIR emission in GX 339--4 during
outburst. In the following sections we discuss how the other
suggested components can explain the observed optical/nIR behavior.

\begin{figure*}
\centerline{\includegraphics[angle=-90,width=8cm]{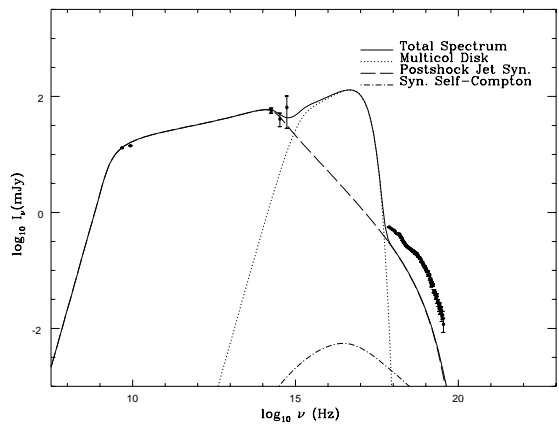}}
\caption{An alternate model for the hard state SED of Fig.~4a,
representing minimal jet contribution.  This is a similar model to
that shown in Fig.~\ref{fig:jetmodel}b, but the base of the jet is
assumed not to radiate, leaving the hard X-rays to originate in
coronal emission (not shown).  Only after acceleration is the jet
plasma radiative in synchrotron (long dashed line) and synchrotron
self-Compton (dot-dashed line) emission.  In this scenario, the I/V
bands must either originate in the corona, or in thermal multicolor
thin disk emission as shown here.  The thermal emission total
luminosity shown here is $\sim 60\%$ $L_{\rm Edd}$ for a 7 $M_\odot$
central object, which may be excessive for even a bright hard state.
One sees that in order to explain the H-band emission with jet
synchrotron, it is quite hard to avoid synchrotron contributing
significantly in the X-rays as well, and this is a conservative model.
In order to reduce this contribution, either the spectral index of the
emission must be very steep, or only $\ll 1\%$ of the particles can be
accelerated. \label{nonoz}}
\end{figure*}

\subsubsection{Hard state}\label{sec:hard_state}

During the rise in the hard state of GX 339--4 we observed a strong
correlation between the nIR and X-ray fluxes.  Their relationship has
a similar power-law form as the radio/X-ray correlations observed in
the hard states of GX 339--4 and other black-hole binaries
\citep{hahuca1998,conofe2003,gafepo2003}, though with a slope of
0.53$\pm$0.02 compared to 0.71$\pm$0.01.  Our slope was
calculated using the full 3--100 keV X-ray band rather than the 2--11
keV band used by \citet{gafepo2003}, but extends over a similar range
in X-ray flux as the radio/X-ray correlations.  While the origin of
the X-ray emission in the hard state is still a matter of debate, the
radio emission is generally believed to be from a jet outflow
\citep[for a recent review see][]{fe2003}. 

The similarity between the nIR/X-ray and radio/X-ray relations
suggests that most of the nIR emission in the hard state could
originate in the jet as well. This was already suggested by
\citet{cofe2002} who found that the nIR emission in the 1981 and 1997
hard states of GX 339--4 lies on an extrapolation of the radio
spectrum of the source. They interpreted the nIR emission as being
optically thin synchrotron emission from a compact jet.  This
interpretation is supported by several additional arguments.  First,
neither thermal emission from a (heated) disk\footnote{Following
\citet{vrraga1990} we calculated the spectrum of an irradiated
accretion disk (outer radius $3\cdot10^{11}$ cm) and found that for a
range of reasonable parameters it peaks at wavelengths well below the
$H$-band. X-ray reprocessing can therefore not explain the observed
upturn between the $I$ and $H$ bands.} or companion star, nor a
magnetically dominated compact corona, can explain the nIR excess
that is apparent from our two hard state SEDs, since these models
predict an decrease toward longer wavelengths in the optical/nIR. 
Synchrotron emission from a compact jet can naturally account for
this excess, as it is predicted to peak in the nIR for X-ray
binaries, and is intrinsically much redder than the other components
\citep{mafafe2001}. Second, a disk origin of the optical/nIR can
probably also be ruled out on the basis of its observed decay by
factors of $\sim$5--15 around MJD 52400. Such changes should occur on
a viscous time scale, which in this system is much longer (probably
15--20 days, as we argue in \S \ref{subsec:spl}). Reprocessing could
in principle explain the fast time scales, since the optical would
almost immediately follow the X-rays. However, if the sudden drop
around MJD 52400 was the result of a decrease in the illuminating
X-ray flux (assuming a constant outer disk geometry) the optical/nIR
spectrum should have become redder (because of the lower outer disk
temperature), while in fact the opposite was observed. Finally, as
soon as the disk black body started to contribute significantly in
the (3--100 keV) X-ray band, the nIR flux decreased dramatically.
This behavior is similar to what is seen for the  radio flux in GX
339--4, which is quenched in the spectrally soft state
\citep{fecotz1999}.

Although the above suggests that the nIR in the hard state is
dominated by emission from a (compact) jet, extended nIR emission
from galactic black hole binaries has so far only been observed once,
in the micro-quasar GRS 1915+105 \citep{saecsu1996}.  The typical
size scales of the IR-emitting regions are in general too small to be
easily observed with worse than subarcsecond resolution.  

We can look to the light curve for some clues about contributions
from other mechanisms. The fact that the optical/nIR emission did not
return to pre-outburst levels after the MJD $\sim$52400 decrease
probably indicates that, during the hard state, other components in
addition to a jet contributed to the optical/nIR.  The steep ``V''
shape of the SED from the nIR to optical also suggests two distinct
components to the emission, and while it seems fairly likely that the
$H$-band data point is dominated by jet synchrotron, the $I$- and
$V$-bands could be combinations of various components.  Possible
contributors to the $I/V$ bands are thermal emission from an
accretion disk or even irradiation of the outer disk or companion
star as suggested by \cite{cofe2002}.  While spectrally direct disk
emission could be possible (see Fig.~\ref{nonoz}), the fast optical
timescales observed are inconsistent with the expected thin disk
viscous timescales and they limit the thermal contribution from
the disk. The maximum disk contribution in the hard state is probably
close to the level observed in the intermediate state after MJD
52400.  What we know is that the slope of the blue side of the ``V''
remained remarkably stable over the rising hard state, while the
lower half seems to be rising more, comparatively. So whatever the
underlying physics, it is clear that these two components are
somewhat decoupled.

Most likely, the rise in the SED above the nIR is due to some
combination of thermal and non-thermal emission, but dominated by
synchrotron/cyclotron emission originating in a different region from
the outer jet responsible for the radio through nIR emission.  This
could either be the foot point, or base, of the jet, or some kind of
magnetized compact corona \citep[e.g.,][]{fagumo1982,medifa2000},
though the difference between these two scenarios may be semantical.
As a means for discussion, in Figs.~\ref{fig:jetmodel} \&
\ref{nonoz}, we show examples of two extreme cases involving jets.
Figure \ref{fig:jetmodel} shows the case when a jet is responsible
for most of the SED of the two observed hard states.  We stress that
these are not fits to the data, because they have not been convolved
with the response matrices for the PCA detector, nor do they include
fine features such as reflection and iron lines.  However, they
represent the maximal jet model contribution to the SED, and can help
define what is possible in a more conservative approach, shown in the
opposite extreme in Fig.~\ref{nonoz}.  

The models in Figure \ref{fig:jetmodel} contain 3 basic components:
the $V$ band is dominated by synchrotron radiation from the base of
the jet, and the same emitting electrons inverse Compton scatter with
these synchrotron photons to create the bump dominating the higher
X-ray frequencies.  This base component must be slightly decoupled
from the rest of the jet, which creates the radio-IR via
self-absorbed synchrotron emission, or else it would be difficult to
explain the steep ``V'' in the SED.  Such a decoupling is not
expected, but could result from extreme cooling or escape of the
radiating particles before a zone of (re-)acceleration, whereas in a
non-decoupled case the particles are predominantly cooling
adiabatically. The models in Fig.~\ref{fig:jetmodel} were calculated
for a high inclination angle of 60 degrees. Varying this angle by
$\pm$10 degrees leads to similar results, whereas angles of lower
than 30--40 degrees would have resulted in flatter spectra and angles
higher than 70--80 degrees (for which there is no evidence) in
steeper spectra and a loss of power. 

We take the opposite approach in Fig.~\ref{nonoz}, and show a model
for the hard state SED of MJD 52368, where the I/V data result mainly
from direct multicolor thermal disk emission (playing somewhat
devil's advocate to show an extreme case despite arguments against
it).  The jet is assumed to radiate only beyond an acceleration zone,
producing optically thick radio to nIR emission, and leaving room for
a separate corona to produce the hard X-rays (not shown). The disk
model shown has fairly extreme parameters, comprising $\sim0.6 L_{\rm
Edd}$ in power for a 7$M_\odot$ black hole, with an inner disk
temperature of only $\sim 0.2$ keV.  This temperature was chosen to
fall just below the soft X-ray data points, as is normally expected
for bright hard states.  These choices result in an inner disk
truncated at $\sim 150 r_g$ and the outer disk ends at $\sim10^5 r_g$
in order to go through the IR/optical data.  A lower temperature
would result in a similar fit for slightly lower luminosities, and
thus greater values for the inner disk radius.   This model is quite
artificial in that the jet model is simply set to zero before the
acceleration zone, and in reality there will be some emission from
the base, or the corona is equivalent to the base of the jet and a
self-consistent relationship to the outer jet must be worked out.

Either class of model has difficulty reproducing the steep ``V'',
which will be a tight constraint for any theory.  Within the context
of an entirely jet-dominated model, a more sharply peaked particle
distribution than quasi-thermal in the base, or modified geometry
could give a better fit to the ``V''.  What is important to note is
that jets have the potential to contribute greatly to the SED beyond
the radio/nIR, and in both cases the jet synchrotron emission is
difficult to suppress.  In order to fit the assumed optically
thick-to-thin turnover near the $H$-band, and be significantly under
the X-rays, either the power-law of accelerated particles must be
very steep, or only a tiny fraction of the total number of particles
can be accelerated.  In these models we assumed $\sim 5-10\%$ of the
particles were accelerated, with particle distribution spectral
indices typical of shock acceleration ($2-2.4$).  Acceleration via
other mechanisms such as magnetic reconnection as seen in solar
flares would result in even harder spectral indices.  

The changes in the two hard state SEDs over time can be accounted for
in jet-dominated models by varying the total power and with slight
changes in the geometry and equipartition of energy.  We plan to
explore these variations along with full fitting to the X-rays in
another paper.

It is not clear if jet emission dominated the nIR (and possibly also
the optical) for the entire hard state. Around MJD 52360 we observed
a strong reddening in the $V-H$ color, possibly indicating that only at
that point the jet started to dominate the optical/nIR emission.
Interestingly, taking only the hard state data after MJD 52360 the
slope of the $H$-band/X-ray flux relation becomes 0.73$\pm$0.03,
which is consistent with the value for the radio/X-ray relation
(0.71$\pm$0.01). It should be noted however, that the slope of the
data before MJD 52360 steepens to 0.65$\pm$0.03, which is also
marginally consistent with 0.71.

\subsubsection{Intermediate state}

The accelerated steepening of the power-law component in the X-ray
spectrum coincides with the source leaving the nIR/X-ray relation
that was observed in the hard state. The steepening of the power-law
component could be the result of increased cooling or less efficient
particle acceleration. If at the same time also the fraction of
accelerated particles goes down, this could explain the sudden drop
in the optical and nIR fluxes. Regardless of the exact interpretation
of the sudden decrease in the optical/nIR, our observations strongly
suggest that the compact jet started to change when the source enters
the intermediate state, before it likely shut off completely in the
spectrally soft state. At the end of the
intermediate state the optical/nIR was probably dominated by the
contribution from the accretion disk (see also \S \ref{subsec:spl}).

After the source entered the intermediate state and then the spectrally soft state, the X-ray power-law component
decreased on a much longer timescale than the optical/nIR component.
This further suggests a different origin for the power-law component
in the other three states, as already mentioned at the end of
\S\ref{sec:comp}.

\subsubsection{Soft state}\label{subsec:spl}

While the optical/nIR light curves in the hard state had a shape
similar to that of the X-ray power-law flux, in the spectrally soft
states they were more similar to that of the X-ray disk flux. The
observed lag (15--20 days) of the X-ray flux variations with respect
to those in the {\it H}-band (Figure \ref{fig:flaring}) strongly
suggest a disk origin of the optical/IR emission in those states. The
magnitude of the lag is in line with what one would expect for the
time it takes for changes in the mass accretion rate in the outer
disk (where most of the optical/nIR originates) to reach the inner
disk (where the X-ray are produced), i.e. the viscous time scale. The
fact that a delay of this order is observed also indicates that X-ray
reprocessing is of minor importance to the nIR continuum, since in
that case changes in the X-rays would have preceded those in the
$H$-band by several seconds. The disk origin of the optical/nIR is
also supported by the changes in the SED. The optical/nIR points of
the MJD 52406 SED were fitted by a power law with slope 1.4$\pm$0.4.
This is marginally consistent with  $F_\nu\propto\nu^2$, which is
expected for the Rayleigh-Jeans part of a multi-color disk spectrum.

The sudden onset of flaring in the $H$-band as the disk component
became the dominant X-ray spectral component (around MJD 52440) is
not well understood. These fast changes are not observed in the X-ray
band. The sampling in the X-rays is arguably not as good as in the
$H$-band but individual X-ray observations made during the
nIR-flaring period show hardly any variability, making a reprocessing
origin unlikely. The flares may have a similar origin as the optical
flares found by \citet{zucash2003} in five quiescent soft X-ray
transients. These have  similar amplitudes, but shorter time scales
(up to hours). \citet{zucash2003} suggested a magnetic loop
reconnection events in the outer accretion disk as a possible
explanation. Rapid optical/IR variability has been observed before in
the hard states of GX 339--4 and XTE J1118+480 \citep{hyhacu2003},
but the time scales involved there are much shorter and the X-rays
always show variations on a similar time scale.

\subsection{X-ray vs.\ optical/nIR delay}

Optical/nIR monitoring campaigns of GRO J1655--40
\citep{orreba1997},  XTE J1550--564 \citep{jabaor2001b}, and 4U
1543--47 \citep{buba2004} revealed that in these sources the start of
the outburst in X-rays lagged the start in the optical/nIR by 3--11
days. These outbursts most likely started in an "off" state
($L_X\sim10^{30-31}$ erg\,s$^{-1}$), and could thus be different from
the event reported here for GX 339-4, which may have started in a
higher luminosity state ($L_X>2\cdot10^{33}$ erg\,s$^{-1}$, based on
the lowest detected flux by \citet{conofe2003} and assuming a
distance of 6 kpc).  Nevertheless, our results may be relevant for
interpreting these other events. The reported delays are generally
thought to reflect the viscous time scale of a thin accretion disk,
which determines the time it takes for the increase in mass accretion
rate to reach the inner, X-ray emitting regions of the accretion
flow, and/or the time it takes to replace a radiatively inefficient
advection dominated accretion flow  \citep[ADAF, see
e.g.][]{halamc1997}. In both the above cases the measurements of the
delays were based on a comparison of optical/nIR with {\it RXTE}/ASM
data, which is far less sensitive than the {\it RXTE}/PCA. Our data
set for the 2002 outburst of GX 339--4 shows that a source may
already be rising for several weeks before the ASM is able to detect
it. The X-ray delay that we measure with the PCA is less than a week
and consistent with zero delay, whereas the ASM yields a delay of
20-45 days, a factor of more than 3--6 longer. Obviously, this has
important consequences for conclusions based on delays measured with
the ASM. For example, if the delays estimated from the ASM light
curves were longer than the true delay by the same factor as found
in  the case of GX 339--4, then the true delays in GRO J1655--40, XTE
J1550--564, and 4U 1543--47 may have been only 0.5--2 days.

Delays shorter than a few days are difficult to understand in disk
instability models, which predict the X-rays (from the inner disk or
ADAF) to rise about a week after the optical/nIR (from the outer
disk). No delays are expected if both the hard X-rays and 
optical/nIR are produced by a jet or the base of a jet. In fact, the
optical/nIR should lag the X-ray in such a case, by seconds to
minutes. However, it is not clear how the jet is fed with matter
during the early stages of the outburst, as there is no sign of an
inner accretion disk in the early X-ray spectra. One option is that
the jet is not fed by the inner accretion disk, but rather by an
independent flow, that is able to respond much faster to changes in
the accretion rate than the accretion disk itself, and possibly
couples directly to the jet. The existence of such a second flow has
recently been discussed by several authors
\citep{va2001,smhesw2002}. 

\subsection{X-ray reflection and reprocessing}

In the parts of the spectrally soft state and the intermediate state
at the end of the 2002/2003 outburst Fe K$\alpha$ emission lines and
reflection components in the X-ray spectrum of GX 339--4
\citep{mifare2004,mirafa2004} provide strong evidence for a source
that irradiates the inner accretion flow with hard X-rays. Assuming a
concave shape for the accretion disk, one would expect that such a
source of hard X-rays might also to be able to irradiate the outer
parts of the accretion disk, which should be visible as enhanced
optical/nIR flux. However, the fact that no clear evidence is found
for reprocessed X-rays in the optical/nIR suggest that the geometry
of the irradiating source is not efficient for X-ray heating of the
outer disk. This would favor a geometry for the hard X-ray source in
which it is aligned with the disk plane (and does not radiate
isotropically) rather than it being elevated above the disk.

\section{Summary \& Conclusions}\label{sec:sum}

Both from a spectral and variability viewpoint it is becoming
increasingly clear that in some X-ray states the optical/nIR emission
in black hole X-ray binaries must have a strong non-thermal component
\citep[see e.g.][]{cofe2002,kastsp2001,hyhacu2003}. Our observations
revealed, in great detail, a switch in the dominating optical/nIR
emission mechanism as GX 339--4 moved from the hard X-ray state to
intermediate state and then the spectrally soft state. In the hard
state we found a correlation between the optical/nIR and X-ray fluxes
that extended over three orders of magnitude in X-ray flux and which
had a similar slope as the one found between radio and X-ray fluxes
in several black hole binaries, including GX 339--4. The
non-measurable delay between X-ray and optical/nIR fluxes and the
spectral energy distribution suggest a non-thermal/jet origin for the
nIR emission in the hard state. The optical emission in the hard
state is probably due to a combination of emission from a jet, a
(heated) accretion disk and possibly a compact corona. As soon as the
source entered the intermediate state we observed a large decrease in
the optical/nIR fluxes, which suggests changes in the jet properties,
even though the radio stayed at a similar level as in the hard state.
In the spectrally soft state the ratio of X-ray to optical/nIR was a
factor of 10 higher than in the hard state and the delay between nIR
and X-ray fluxes was more than two weeks. These changes in the X-ray
and optical/nIR properties suggest that the dominant source of
optical/nIR emission was a non-heated (or only moderately heated)
accretion disk. It is interesting to note that in none of the X-ray
states X-ray reprocessing in the outer disk seems to be the dominant
source of optical/nIR emission.

Our monitoring observations have shown that the optical/nIR
properties of black hole X-ray binaries are extremely sensitive to
the X-ray state of the system. Especially the X-ray to optical/nIR
flux ratio may be helpful in determining the state of systems for
which accurate X-ray spectral and variability properties cannot be
obtained. 

 Finally, we want to stress the importance of extending optical/nIR
monitoring of black hole transients to perform  variability studies
simultaneously with X-ray observations. Optical variability studies
have often been done without simultaneous X-ray coverage
\citep{imkrmi1990,stimmi1990,stscim1997} and in cases were X-ray data
was obtained, the observations were predominantly in the hard X-ray
state \citep{kastsp2001,hyhacu2003}. Similar observations in other
X-ray states will greatly increase understanding of the accretion
flows in these systems, as the changes in the optical/nIR variability
properties (e.g. energy dependence over many orders of QPO frequency)
can be related to changes in the spectral energy distribution and be
used to rule out possible sources of optical/nIR emission. For
example, recent X-ray and optical observations of the black hole
transient H1743--322 by \citet{spstka2004} seem to indicate that
so-called type B X-ray QPOs (observed in parts of the spectrally soft
states) no optical counterpart unlike those in the hard and
intermediate states (the so-called type C QPOs), supporting claims
that the optical/nIR in these two states have very different origins.

\acknowledgments

The authors would like to thank Suzanne Tourtellotte in assisting
with the optical/nIR data reduction, David Gonzalez Huerta and Juan
Espinoza for observing at CTIO, Rob Fender and Jon Miller for useful
suggestions and comments, and finally Elena Gallo and Stephane Corbel
for re-analyzing their radio observations.  J.\ Homan gratefully
acknowledges support from NASA.  M.\ Buxton and C.D. Bailyn are
supported through the NSF grant AST-0098421. S. Markoff is supported
by an NSF Astronomy \& Astrophysics postdoctoral fellowship under NSF
Award AST-0201597.

\end{document}